\documentclass[aps,prb,twocolumn,groupedaddress,showpacs]{revtex4}
\usepackage{epsfig}
\usepackage{color}
\begin{document}

\title{Magnetic and Superfluid Transitions in the
d=1 Spin-1 Boson Hubbard Model}
\author{G.G. Batrouni$^1$, V.G. Rousseau$^2$, and R.T. Scalettar$^3$} 
\affiliation{
$^1$INLN, Universit\'e de Nice-Sophia Antipolis, CNRS; 
1361 route des Lucioles, 06560 Valbonne, France
}
\affiliation{
$^2$Instituut-Lorentz, LION, Universiteit Leiden, Postbus 9504, 2300 RA
Leiden, The Netherlands}
\affiliation{
$^3$Physics Department, University of California, Davis, CA 95616
}

\begin{abstract}
The interplay between magnetism and metal-insulator transitions is
fundamental to the rich physics of the single band fermion Hubbard
model.  Recent progress in experiments on trapped ultra-cold atoms
have made possible the exploration of similar effects in the boson
Hubbard model (BHM).  We report on Quantum Monte Carlo (QMC)
simulations of the spin-1 BHM in the ground state.  For
antiferromagnetic interactions, $(U_2>0)$, which favor singlet
formation within the Mott insulator lobes, we present exact numerical
evidence that the superfluid-insulator phase transition is first
(second) order depending on whether the Mott lobe is even
(odd). Inside even Mott lobes, we examine the possibility of
nematic-to-singlet first order transitions. In the ferromagnetic case
$(U_2<0)$, the transitions are all continuous.  We map the phase
diagram for $U_2<0$ and demonstrate the existence of the ferromagnetic
superfluid. We also compare the QMC phase diagram with a third order
perturbation calculation.
\end{abstract}


\maketitle

The single band fermion Hubbard model (FHM) offers one of the most
fundamental descriptions of the physics of strongly correlated
electrons in the solid state.  The spinful nature of the fermions is
central to the wide range of phenomena it displays such as interplay
between its magnetic and transport properties.\cite{fazekas} Such
complex interplay is absent in the superfluid to Mott insulator
transition \cite{fisher89,batrouni90,freericks96,prokofev07} in the
spin-0 Boson Hubbard model (BHM).  However, purely optical traps
\cite{stamperkurn98} can now confine alkali atoms $^{23}$Na, $^{39}$K,
and $^{87}$Rb, which have hyperfine spin $F=1$, without freezing
$F_z$.  As in the fermion case, the nature of the superfluid-Mott
insulator (SF-MI) transition is modified by the spin fluctuations
which are now allowed.  Initial theoretical work employed continuum,
effective low-energy Hamiltonians and determined the magnetic
properties and excitations of the superfluid phases.\cite{magprop}

To capture the SF-MI transition
it is necessary to consider the spin-1 Bosonic Hubbard Hamiltonian,
\begin{eqnarray} 
\label{hubham} H=&& -t \sum_{\langle ij\rangle
,\sigma}(a^{\dagger}_{i\sigma}a_{j\sigma}+ h.c) + \frac{U_0}{2} \sum_i
{\hat{n}_i}({\hat{n}_i}-1) \nonumber \\ &&+\frac{U_2}{2} \sum_i
(\vec{F}_{i}^{\, 2}- 2 \, \hat{n}_i)
\end{eqnarray} 
The boson creation (destruction) operators $a_{i \sigma}^\dagger
\,(a_{i \sigma})$ have site $i$ and spin $\sigma$ indices.
$\sigma=1,0,-1$.  The first term describes nearest-neighbor, $ \langle
ij \rangle$, jumps. The hybridization $t=1$ sets the energy scale, and
we study the one dimensional case.  The number operator $\hat
n_i\equiv\sum_\sigma\hat{n}_{i\sigma}= \sum_\sigma
a^{\dagger}_{i\sigma}a_{i\sigma}$ counts the total boson density on
site $i$.  The on-site repulsion $U_0$ favors states with uniform
occupation and competition between $U_0$ and $t$ drives the MI-SF
transition.  The spin operator $\vec{F}_i=\sum_{\sigma,\sigma '}
a^{\dagger}_{i\sigma} \vec{F}_{\sigma \sigma '} a_{i\sigma '}$, with $
\vec{F}_{\sigma \sigma '}$ the standard spin-1 matrices, contains
further density-density interactions and also interconversion terms
between the spin species. We treat the system in the canonical
ensemble where the total particle number is fixed and the chemical
potential is calculated, $\mu(N)=E(N+1)-E(N)$ where $E(N)$ is the
ground state energy with $N$ particles. This holds only when each
energy used is that of a single thermodynamic phase and not that of a
mixture of coexisting phases as happens with first order transitions.
It is therefore incorrect to determine first order phase boundaries
with the naive use of this method. In the present case there is the
added subtlety that the three species are interconvertible. These
issues will be addressed in more detail below.

Several important aspects of the spin-1 BHM are revealed by analysing
the independent site limit, $t/U_0=0$.  The Mott-1 state with $n_i=1$
on each site has ${\cal E}_{\rm M}(1)=0$.  In the Mott-2 state with
$n_i=2$, the energy is ${\cal E}_{\rm M}(2)=U_0-2U_2$, if the bosons
form a singlet, $F=0$, and is ${\cal E}_{{\rm M}2}=U_0+U_2$ if $F=2$.
Thus $U_2>0$ favors singlet phases while $U_2<0$ favors (on-site)
ferromagnetism.  This applies to all higher lobes as well. In the
canonical ensemble, the chemical potential at which the system goes
from the $n$th to the $(n+1)$th Mott lobe is $\mu(n\to n+1)={\cal
  E}_{n+1}-{\cal E}_{n}$.

First consider $U_2>0$. The energy of Mott lobes at odd filling,
$n_{\rm o}$, is ${\cal E}_{M}(n_{\rm o})=U_0n_{\rm o}(n_{\rm
  o}-1)/2+U_2(1-n_{\rm o})$ while at even filling, $n_{\rm e}$, ${\cal
  E}_{M}(n_{\rm e})=U_0n_{\rm e}(n_{\rm e}-1)/2-n_{\rm
  e}U_2$. Therefore, the boundaries of the lobes, going from lower to
higher filling, are $\mu (n_{\rm e}\to (n_{\rm e}+1))=n_{\rm e}U_0$
and $\mu (n_{\rm o}\to (n_{\rm o}+1))=n_{\rm o}U_0-2U_2$. This demarks
the positions of the `bases' of the Mott lobes in the
$(t/U_0,\mu/U_0)$ ground state phase diagram.  For $U_2>0$ the even
Mott lobes grow at the expense of the odd ones, which disappear
entirely for $U_0=2U_2$.

For $U_2<0$, the ground state is ferromagnetic (maximal $F$:
$F^2=n(n+1)$) which gives for all Mott lobes ${\cal E}_{\rm
  M}(n)=n(n-1)(U_0+U_2)/2$. Consequently, the boundary of the $n$th
and $(n+1)$th Mott lobes $\mu (n\to n+1)=n(U_0+U_2)$.  The bases of
both the odd and even Mott lobes shrink with increasing $|U_2|$, in
contrast to the $U_2>0$ case where the even Mott lobes expand.

Mean-field treatments of the lattice model capture the SF-MI
transition as the hopping $t$ is turned on, and have been performed
both at zero and finite
temperature.\cite{krutitsky04,tsuchiya05,pai08} Even when $U_2=0$, the
spin degeneracy alters the nature of the transition.  For $U_2>0$, the
order of the phase transition depends on whether the Mott lobe is even
or odd. These mean field calculations assume a non-zero order
parameter $\langle a_{i\sigma} \rangle$, which cannot be appropriate
in d=1 or in d=2 at finite $T$.  Therefore it is important to verify
these predictions for the qualitative aspects of the phase diagram,
especially in low dimension.  A quantitative determination of the
phase boundaries requires numerical treatments.  Indeed, DMRG
\cite{dmrg} and Quantum Monte Carlo (QMC) \cite{apaja06} results for
$U_2>0$ and d=1 reported the critical coupling strength and showed the
odd Mott lobes are characterized by a dimerized phase which breaks
translation symmetry.

For $U_2<0$, the nature of the SF-MI transition does not depend on the
order of the Mott lobe while for $U_2>0$ it is predicted to be
continuous (discontinuous) into odd (even) lobes. Consequently, it
suffices to study the first two Mott lobes both for $U_2>0$ and
$U_2<0$ to demonstrate the behavior for all lobes. Furthermore, in
what follows we will focus on the case $|U_2/U_0|=0.1$ in order to
compare our results for $U_2>0$ with Rizzi {\it et al}\cite{dmrg}

Here, we will use an exact QMC approach, the SGF algorithm with
directed update, to study the spin-1 BHM in d=1 for both positive and
negative $U_2$.\cite{SGF}

\begin{figure}
\epsfig{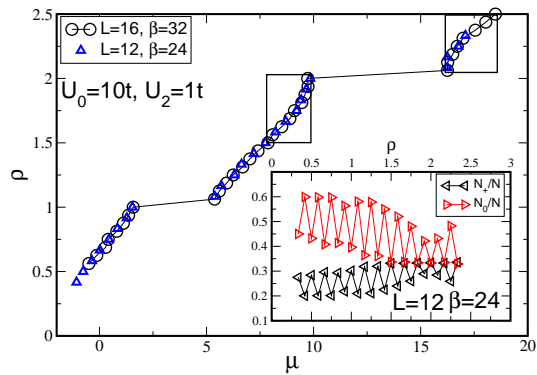}
\caption{(Color online) $\rho(\mu)$ exhibits Mott plateaux: gapped,
  insulating phases at commensurate fillings. Inset: $N_+/N$ and
  $N_0/N$ {\it vs} $\rho=N/L$. Singlets form where $N_+/N$ and $N_0/N$
  are equal (see text).  }
\label{rhovsmuU010Ut1fss}
\end{figure}

For $U_2>0$ (e.g.~$^{23}$Na), which favors low total spin states,
Fig.~\ref{rhovsmuU010Ut1fss} shows the total number density $\rho=N/L$
against the chemical potential, $\mu$, for $U_0=10t$ and $U_2=t$. It
displays clearly the first two incompressible MI phases.  In agreement
with the $t/U_0=0$ analysis, $U_2$ causes an expansion of the second
Mott lobe, $\rho=2$, at the expense of the first, $\rho=1$. Our Mott
gaps agree with DMRG results\cite{dmrg} to within symbol size.
However, the $\rho$ versus $\mu$ curve in Fig.~\ref{rhovsmuU010Ut1fss}
does not betray any evidence of the different natures of the phase
transitions into the first and second Mott lobes. In particular, for a
spin-0 BHM, first order transitions are clearly exposed by the
appearence of negative compressibility,\cite{batrouni00}
$\kappa=\partial \rho/\partial\mu$, which is not present here. The
transition into the second Mott lobe is expected to be first order and
driven by the formation of bound pairs of bosons in singlet
states. Therefore, in the canonical ensemble, we expect that near this
transition there will be phase coexistence between singlets arranged
in a Mott region and superfluid.

The nature of the transitions is revealed by the evolution of the spin
populations in the system as $\rho$ increases. Since the singlet
wavefunction is
$|0,0\rangle=\sqrt{2/3}|1,1\rangle|1,-1\rangle-\sqrt{1/3}|1,0\rangle|1,0\rangle$,
this state has $\rho_+=\rho_0=\rho_-$. We plot in the inset of
Fig.~\ref{rhovsmuU010Ut1fss} the population fractions, $N_0/N$ and
$N_-/N=N_+/N$ versus the total density. We see that as $\rho$
increases, $N_+/N$ and $N_0/N$ oscillate: When $N$ is even, singlet
bound states of two particles try to form drawing the values of
$N_+/N$ and $N_0/N$ closer together. However, singlets form fully,
making $N_+=N_0$, only close to the second Mott lobe, $\rho=2$, where
we clearly see $N_+/N=N_0/N$ for a range of even values of $N$. On the
other hand, when $N$ is odd, singlets cannot form and the spin
populations are much farther from that given by the singlet
wavefunction. In the thermodynamic limit and fixed $N$, one expects
true phase separation into $\rho=2$ singlet MI regions and $\rho<2$ SF
regions. For a finite system, phase separation commences for (even)
fillings where we first have $N_+/N=N_0/N$, {\it i.e.}  $1.5\leq \rho
< 2$ and similar behavior for $\rho > 2$. Another interesting feature
in the inset of Fig.~\ref{rhovsmuU010Ut1fss} is that the difference
between $N_+/N$ and $N_0/N$, for even $N$, decreases linearly as the
density approaches the transition at $\rho= 1.5$. No such behavior is
seen as the first Mott lobe is entered from below or above $\rho=1$:
the transition is coninuous as predicted for odd lobes.

The boxes in Fig.~\ref{rhovsmuU010Ut1fss} show the values of $\rho$
corresponding to phase coexistence rather than to one stable
thermodynamic phase. It is, therefore, clear that the canonical
calculation of the phase boundaries, {\it i.e.} simply adding a
particle to, or removing it from, the MI, is not applicable in the
presence of a first order transition.

\begin{figure}
\epsfig{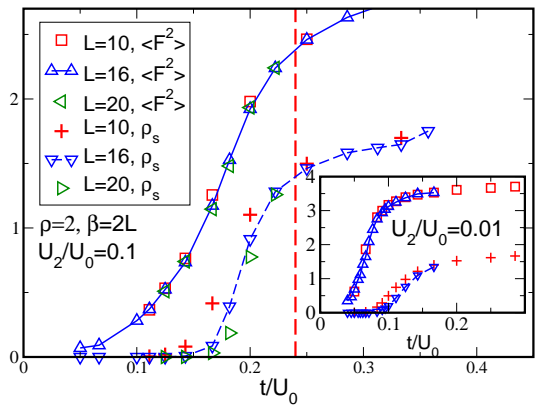}
\caption{(Color online) The square of the local moment, $\langle F^2
  \rangle$, and the superfluid density, $\rho_s$, vs $t/U_0$ for
  $\rho=2$.  In the Mott lobe, $\langle F^2 \rangle\to 0$ signaling
  singlet formation and $\rho_s \to 0$ indicating an insulator. The
  dashed line indicates the critical $t/U_0$ from Rizzi {\it et
  al}\cite{dmrg}. Inset: Same but with $U_2/U_0=0.01$.}
\label{jsqvsU}
\end{figure}

Figure~\ref{rhovsmuU010Ut1fss} reveals the SF-MI transition at fixed,
sufficiently large $U_0$ when $\rho$ is varied. In Fig.~\ref{jsqvsU}
we show the transition when the density is fixed, $\rho=2$, and $U_0$
is the control parameter with $U_2/U_0$ fixed at $0.1$ (main figure)
and $0.01$ (inset). Singlet formation is clearly shown by $\langle
F^2\rangle\to 0$, as $t/U_0$ decreases and the second MI is
entered.\cite{footnote1} Indeed, the origin of the first order
transition into even Mott lobes, as the filling is tuned, is linked to
the additional stabilization of the Mott lobe associated with this
singlet energy.\cite{pai08} The superfluid density, $\rho_s=L\langle
W^2\rangle/2t\beta$, where $W$ is the winding number, is a topological
quantity and truly characterizes the SF-MI phase transition which is
continuous in this case. As $L$ is increased from $10$ to $16$ and
$20$, the vanishing of $\rho_s$ gets sharper.  We find that the
critical value of $t/U_0$ for the $\rho=2$ lobe is somewhat less than
that reported in DMRG\cite{dmrg} indicated by the dashed line. We
believe this is because with DMRG the phase boundaries were obtained
using finite differences of the energy with small doping above and
below commensurate filling.  As discussed above, this is not
appropriate for a first order transition.

For $2dU_2/U_0<0.1$, and $d=2,3$, mean field\cite{zhou,imambekov04}
predicts that, when $t/U^{c1}_0\sim \sqrt{U_2/4dU_0}$ is in the MI,
then Mott lobes of even order are comprised of two phases: (a) the
singlet phase for $t/U_0\leq t/U^{c1}_0$ and (b) a nematic phase for
$t/U^{c1}_0 \leq t/U_0\leq t/U^c_0$, where $t/U^c_0$ is the tip of the
Mott lobe. Inside the lobe, the nematic-to-singlet transition is
predicted to be first order which raises the question: are the
singlet-to-SF and the nematic-to-SF transitions of the same order?
Figure~\ref{jsqvsU} shows that the SF-MI transition, $\rho_s\to 0$,
occurs at larger $t/U_0$ than singlet formation, $\langle F^2\rangle
\to 0$, both for $U_2/U_0=0.1$ and $0.01$. The passage of $\langle
F^2\rangle$ to zero gets sharper for smaller $U_2/U_0$ but remains
continuous, not exhibiting any signs of a first order transition. We
have verified this for $U_2/U_0=0.1, \,0.05,\, 0.01,
\,0.005$. Furthermore, the insensitivity of $\langle F^2\rangle $ to
finite size effects indicates that it does not undergo a continuous
phase transition. We have also verified that for the second Mott lobe,
the SF-MI transition is first order regardless whether $t/U_0$ is less
or greater than $t/U^{c1}_0$.  We conclude that while $\rho_s\to 0$ is
a continuous critical transition, $\langle F^2\rangle \to 0$ is a {\it
  crossover} not a phase transition. This, of course, does not
preclude the possibility of a first order transition for $d=2,3$.

\begin{figure}
\epsfig{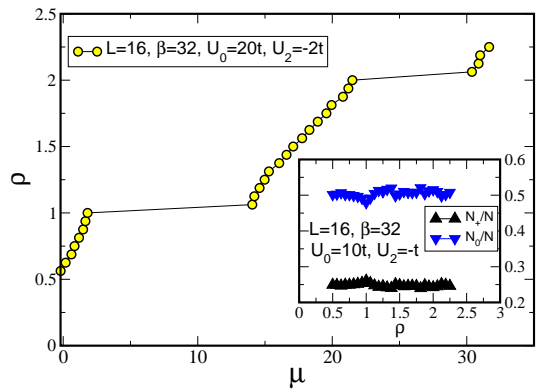}
\caption{(Color online) The density, $\rho$ versus chemical potential,
  $\mu$, exhibits the usual Mott plateaux at commensurate filling for
  $U_2 < 0$. Inset: The spin population fractions showing
  $\rho_0=2\rho_+$. They do not exhibit oscillations like in
  Fig.~\ref{rhovsmuU010Ut1fss}. }
\label{rhovsmuU010Utm1}
\end{figure}

Whereas $^{23}$Na has positive $U_2$, $^{87}$Rb has $U_2<0$, leading
to different behavior.  We begin with $\rho$ versus $\mu$ in
Fig.~\ref{rhovsmuU010Utm1}.  Unlike the $U_2>0$ case, the SF-MI
transitions are continuous for both even and odd Mott lobes: the inset
shows that the spin populations do not oscillate as for $U_2>0$. The
population ratio, $\rho_0=2\rho_+$ can be understood as follows. As
shown above for $t/U_0\to 0$, maximum spin states are favored when
$U_2<0$. So, when a site is doubly occupied, the spin-2 state is
favored. But, since our study is in the $S_z^{\rm total}=N_+-N_-=0$
sector, the wavefunction of the spin-2 state is $|2,0\rangle=
1/\sqrt{3}|1,1\rangle|1,-1\rangle + \sqrt{2/3}|1,0\rangle|1,0\rangle$
and thus $\rho_0=2\rho_+=2\rho_-$.

\begin{figure}
\epsfig{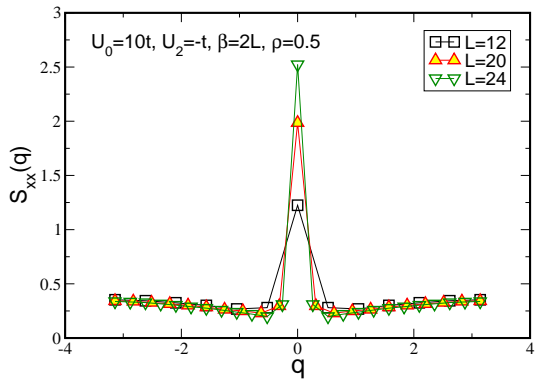}
\caption{(Color online) The magnetic structure factor $S_{\rm xx}(q)$
  for $U_2<0$ and $\rho=0.5$ exhibits a sharp $q=0$ peak indicating
  ferromagnetic order.  }
\label{corrmagxFTutm1}
\end{figure}

As discussed above, $U_2<0$ favors `local ferromagnetism', namely high
spin states on each of the individual lattice sites.  As with the FHM,
the kinetic energy gives rise to second order splitting which lifts
the degeneracy between commensurate filling strong coupling states
with different intersite spin arrangements.  We can therefore ask
whether the local moments order from site to site: Do the Mott and
superfluid phases exhibit global ferromagnetism\cite{pai08}?  To this
end, we measure the magnetic structure factor,
\begin{equation} 
S_{\rm \sigma \sigma}(q) = \sum_{l} e^{iql} \langle F_{\sigma,j+l}
F_{\sigma,j} \rangle 
\end{equation} 
where $\sigma=x$ or $z$.  Figure~\ref{corrmagxFTutm1} shows $S_{\rm
  xx}(q)$ in the superfluid phase at half-filling.\cite{footnote2} The
peak at $q=0$ grows linearly with lattice size, indicating the
superfluid phase does indeed possess long range ferromagnetic order.
We find that the MI phase is also ferromagnetic.

\begin{figure}
\epsfig{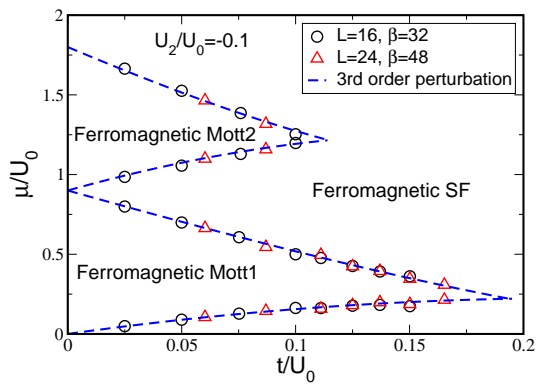}
\caption{(Color online) Phase diagram of the spin-1 BHM for
  $U_2/U_0=-0.1$.}
\label{MottLobesU2neg}
\end{figure}

To determine the phase diagram, we scan the density as in
Fig.~\ref{rhovsmuU010Utm1} for many values of $U_0$ with $U_2/U_0$
constant ($-0.1$ in our case). The resulting phase diagram is shown in
Fig.~\ref{MottLobesU2neg}.  Comparison of data for two lattice sizes
demonstrates that finite size effects are small.

Early in the evaluation of the phase boundaries of the spin-0 BHM it
was observed that a perturbation calculation \cite{freericks96} agreed
remarkably well with QMC results.\cite{batrouni90} We now generalize
the spin-0 perturbation theory to spin-1 and show a similar level of
agreement with the QMC results.  If we assume the system always to be
perfectly magnetized, then $n$ bosons on a site will yield the largest
possible spin, $F^2=n(n+1)$. Consequently, the interaction term in the
Hamiltonian, Eq.(\ref{hubham}), reduces to $(U_0-U_2)\sum_i
{\hat{n}_i}({\hat{n}_i}-1)/2$, giving a Hamiltonian identical to the
spin-0 BHM but with the interaction shifted to $(U_0-U_2)/2$. One can
then repeat the perturbation expansion to third order in $t/(U_0-U_2)$
to determine the phase diagram.\cite{freericks96} The result is shown
as the dashed line in Fig.~\ref{MottLobesU2neg} and is seen to be in
excellent agreement with QMC.  The agreement further suggests that the
finite lattice effects in the phase diagram are small.  Such a
perturbation calculation is not possible for the $U_0>0$ case since
$F^2$ depends on the phase, SF vs MI, and on the order of the MI lobe.

The (dipolar) interactions between spinful bosonic atoms confined to a
{\it single} trap have been shown to give rise to fascinating ``spin
textures".\cite{stamperkurn06} An additional optical lattice causes a
further enhancement of interactions, and opens the prospect for the
observation of the rich behavior associated with Mott and magnetic
transitions, and comparisons with analogous properties of strongly
correlated solids.\cite{zhou,imambekov04} Here, we have quantified
these phenomena in the one-dimensional spin-1 BHM with exact QMC
methods. We have shown that, for $U_2>0$, the MI phase is
characterized by singlet formation clearly seen for even Mott lobes
where $\langle F^2\rangle \to 0$ as $U_0$ increases. We also showed
that the transition into odd lobes is continuous while that into even
lobes is discontinuous (first order).  We emphasized that the naive
canonical determination of the phase boundaries is not appropriate for
a first order transition. For $U_2<0$, we showed that all MI-SF
transitions are continuous and that both the SF and MI phases are
ferromagnetic.  The phase diagram in the $(\mu/U_0, t/U_0)$ plane
obtained by QMC can be described very accurately using third order
perturbation theory.

Acknowledgements: G.G.B. is supported by the CNRS (France) PICS 3659,
V.G.R. by the research program of the `Stichting voor Fundamenteel
Onderzoek der Materie (FOM)' and R.T.S. by ARO Award W911NF0710576
with funds from the DARPA OLE Program.  We would like to thank
T.B.~Bopper for useful input.


\begin{thebibliography}{99}
\bibitem{footnote1} While $\langle F^2 \rangle$ vanishes for the even
  Mott lobes, for the odd lobes $\langle F^2 \rangle \rightarrow
  2$. This is obvious for $\rho=1$ since there is a single spin-1
  boson on each site.  For $\rho=3$, for example, two of the bosons
  pair into a singlet, again leaving an effective spin-1 on each site.
\bibitem{footnote2} In a simulation in the canonical ensemble in the
  $F_z=0$ sector, $S_{\rm zz}(q=0)$ is constrained to vanish at $q=0$.
  However, its $q \rightarrow 0$ limit gives the $q=0$ values of the
  unconstrained situation.\cite{batrouni90}
\end{thebibliography}
\end{document}